# Method to Determine the Closed-Loop Precision of Resonant Sensors from Open-Loop Measurements

Tomás Manzaneque, Peter G. Steeneken, Farbod Alijani and Murali K. Ghatkesar

***Abstract*—Resonant sensors determine a sensed parameter by measuring the resonance frequency of a resonator. For fast continuous sensing, it is desirable to operate resonant sensors in a closed-loop configuration, where a feedback loop ensures that the resonator is always actuated near its resonance frequency, so that the precision is maximized even in the presence of drifts or fluctuations of the resonance frequency. However, in a closed-loop configuration, the precision is not only determined by the resonator itself, but also by the feedback loop, even if the feedback circuit is noiseless. Therefore, to characterize the intrinsic precision of resonant sensors, the open-loop configuration is often employed. To link these measurements to the actual closed-loop performance of the resonator, it is desirable to have a relation that determines the closed-loop precision of the resonator from open-loop characterisation data.**

**In this work, we present a methodology to estimate the closed-loop resonant sensor precision by relying only on an open-loop characterization of the resonator. The procedure is beneficial for fast performance estimation and benchmarking of resonant sensors, because it does not require actual closed-loop sensor operation, thus being independent on the particular implementation of the feedback loop. We validate the methodology experimentally by determining the closed-loop precision of a mechanical resonator from an open-loop measurement and comparing this to an actual closed-loop measurement.**

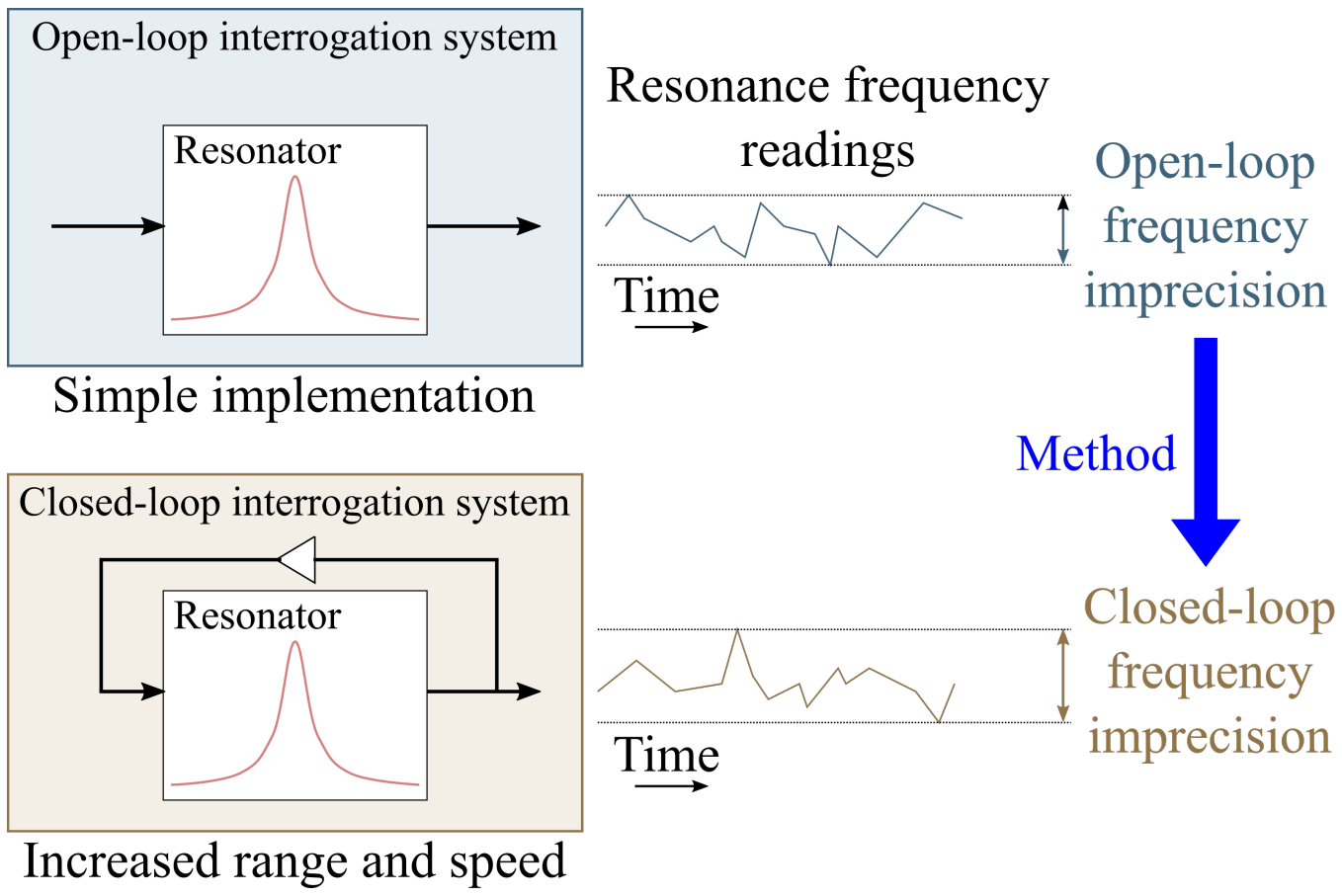

***Index Terms*—Allan deviation, Frequency precision, limit of detection, mass resolution, noise, phase-locked loop, resonant sensor, resonator.**

## I. Introduction

During the last decades, the development of resonant mechanical sensors has been driven by the need for characterizing mass [1]–[3] and stiffness [4]–[6] of micro- and nano-particles with high precision. In resonant mass sensors, the mass of added particles results in a resonance frequency shift, and the minimum particle mass that can be measured is referred to as the mass resolution or limit of detection. For a mass change $\Delta m$ to be resolved via a resonance frequency shift $\Delta\omega_n$, this frequency shift needs to be significant compared to the resonance frequency imprecision $\delta_\omega$, which can be defined as the Allan deviation of a set of measurements of the resonance frequency $\omega_n$, under constant experimental conditions and mass [7]. At a frequency shift equal to the frequency imprecision ($\Delta\omega_n = \delta_\omega$), the mass resolution, imprecision or limit of detection is defined as $\delta_m \equiv \delta_\omega(\mathrm{d}\omega_n\,/\,\mathrm{d}m)^{-1}$, and is found to depend on the frequency imprecision $\delta_\omega$ and the responsivity of the resonance frequency to the mass, $\mathrm{d}\omega_n\,/\,\mathrm{d}m$. Since the responsivity of a mechanical resonator is inversely proportional to the resonator's effective mass $m$, a common approach for improving the limit of detection has been through miniaturizing resonators. It has been shown that microresonators [8], [9] and nanoresonators [10] can reach attogram and zeptogram mass resolution, respectively. By harnessing the vibrations of suspended carbon nanotubes, mass resolution down to yoctogram level has also been demonstrated [11]–[13].

However, reduction of the resonator mass has the drawback that it tends to increase noise-induced frequency imprecision [14], and the consequent miniaturization reduces the effective

This project has received funding from the European Union's Horizon 2020 research and innovation programme under the Marie Skłodowska-Curie grant agreement No 707404, and under Grant Agreement No. 785219 Graphene Flagship. FA acknowledges financial support from European Research Council (ERC) Starting Grant Number 802093. The opinions expressed in this document reflect only the authors' view. The European Commission is not responsible for any use that may be made of the information it contains.

The authors are with the Department of Precision and Microsystems Engineering at Delft University of Technology, Delft, The Netherlands (e-mail: t.manzanequegarcia@tudelft.nl, m.k.ghatkesar@tudelft.nl).

sensor area. Therefore, besides maximizing responsivity, minimizing frequency imprecision $\delta_\omega$ is also of great importance for optimizing the detection limit of resonant mass sensors. The question of what the lowest achievable frequency imprecision is, was first addressed in the context of frequency standards based on thickness-shear-mode quartz resonators [15], [16]. Later on, the problem was studied for resonant sensors based on micro beams [14], [17]. In these works, different sources of imprecision are analysed, that can be classified in three categories: thermomechanical noise, random resonance frequency fluctuations, and instrumentation noise. To analyze the effect of these sources of imprecision on the final mass resolution, it is important to distinguish two different cases: open-loop operation, and closed-loop operation (see the abstract figure).

The open-loop method is rather straightforward to implement, only needing an experimental setup to actuate the resonator and measure its response. There are two typical open-loop resonant sensor readout configurations. In one configuration, the driving frequency is swept across a range around the resonance, and the resonance frequency is determined as the frequency at which the magnitude of the displacement peaks, or at which its phase is shifted by $-\pi/2$ with respect to the driving force. The main drawback of this configuration is that it is rather slow, since measurements at many different frequencies need to be performed. In the second open-loop readout method, the driving frequency is kept constant at or near the resonance frequency, and the phase difference between the response and the driving signal is directly related to the resonance frequency shift. The drawback of this method for practical applications is that the bandwidth and measurement range is limited, even more than in the first open-loop configuration, because the precision of this method rapidly decays if the resonance frequency drifts away significantly more than the peak width, $\omega_n/Q$, from the driving frequency.

To increase the measurement range and speed, feedback (i.e. a closed-loop scheme) is needed to continually adjust the driving frequency, such that it stays near the resonance frequency. Two types of closed-loop schemes for driving resonant sensors can be found in the literature: direct feedback oscillators [18], [19], and phase-locked loops (PLL) [20], [21]. The former need automatic gain controllers to avoid non-linear behaviour at large amplitudes that would be detrimental for the frequency precision, or even more sophisticated controllers to operate the resonator at optimal points of the non-linear regime [22], [23]. The latter rely on phase detection to control the driving frequency, requiring fine tuning the phase detection bandwidth and the controller parameters [21], [24]. Despite requiring extra complexity in the implementation, closed-loop schemes are preferred in most practical sensing scenarios, where significant resonance frequency drifts occur over time, and/or a high number of sensing events (e.g. mass additions) need to be detected per unit time.

The importance of further analyzing precision in open-loop and closed-loop systems is twofold. On the one hand, the theoretical fundamental limits of precision are not the same in both cases, as demonstrated in [21]. On the other hand, when experimentally determining the frequency imprecision of a resonator, different results are obtained from open-loop and closed-loop schemes which are hard to compare. Characterization methods based on open loop are desirable for benchmarking resonators in a simple and standarizable way, since they are rather immediate and do not rely on a particular controller implementation. At the same time, as discussed above, the closed-loop frequency imprecision is the important one in most practical scenarios. Then, the question is: is it possible to infer the closed-loop imprecision from an open-loop measurement? To answer this question, we analyze the transfer functions that govern the conversion of the different noise sources to the final frequency imprecision of the resonant sensor, both in open loop and closed loop. From this analysis, we derive a method to extract the closed-loop imprecision from an open-loop measurement of the resonator. The proposed method captures the contributions of the different potential noise sources in the system and extends beyond earlier studies [14], by also providing valid imprecision estimates for integration times shorter than the resonator's open-loop settling time $\tau_c = 2Q/\omega_n$. To test the applicability, an experimental validation of the method is performed, using an atomic force microscopy (AFM) cantilever as the test device.

## II. Fundamentals

A linear single-degree-of-freedom resonator is fully defined by its transfer function[1] $H(s)$, with effective mass $m$, stiffness $k$, and damping $c$ constants:

$$H(s) = \frac{X(s)}{F(s)} = \frac{1}{ms^2 + cs + k}. \tag{1}$$

$X(s)$ represents the displacement of the resonator from its equilibrium position, and $F(s)$ represents the sum of external forces. The angular resonance frequency is $\omega_n = \sqrt{k/m}$, with resonance frequency $f_n = \omega_n/(2\pi)$, and quality factor $Q = \sqrt{km}/c$.

In this work, resonant sensors that determine mass from the mechanical resonance frequency $\omega_n$ are used as an exemplary case. However, the presented analysis is applicable to other types of resonant sensors too, where for example the sensing functionality is based on changes in the effective stiffness $k$, or where the resonance is not of a mechanical nature, like in electrical $LRC$ resonators, optical or electromagnetic cavity resonators.

For resonant sensors, the resonance frequency must be modelled as a time-dependent parameter, see Fig. 1(a), which results in a parametric variation of the transfer function, like $H(s, \omega_n(t))$. For sensor readout, a measurement scheme is needed to determine the instantaneous value of the resonance frequency. If a harmonic force with constant frequency $\omega_a$ is applied, it follows from (1) that, after a certain settling time of the order of the resonator's characteristic time $\tau_c = 2Q/\omega_n$, a harmonic displacement $x(t) = x_0 \sin(\omega_a t - \varphi_0)$ at the same frequency is obtained, with amplitude $x_0$ and a phase lag $\varphi_0$ with respect to the force. Since this measurement scheme is applied at a constant driving frequency $\omega_a$, in the absence of feedback, the response is referred to as open-loop. When the resonator is driven near resonance ($\omega_a \approx \omega_n$), the resonance frequency can be obtained from the phase of the displacement, $\varphi_0$, by the steady-state relation $\tan\varphi_0 = \omega_a\omega_n/(Q(\omega_a^2 - \omega_n^2))$. For small frequency shifts $\Delta\omega = \omega_a - \omega_n$, a first-order Taylor expansion can be used to obtain a linear relation between $\varphi_0$ and $\Delta\omega$ that can be used for the determination of $\omega_n$ from the measured phase:

$$\varphi_0 = -\frac{\pi}{2} - \frac{2Q}{\omega_n}\Delta\omega. \qquad (2)$$

[1]Throughout the paper, capital letters, like $X(s)$ with complex frequency $s$, are used to denote the Laplace transform of the corresponding lower case time-domain function, like $x(t)$.

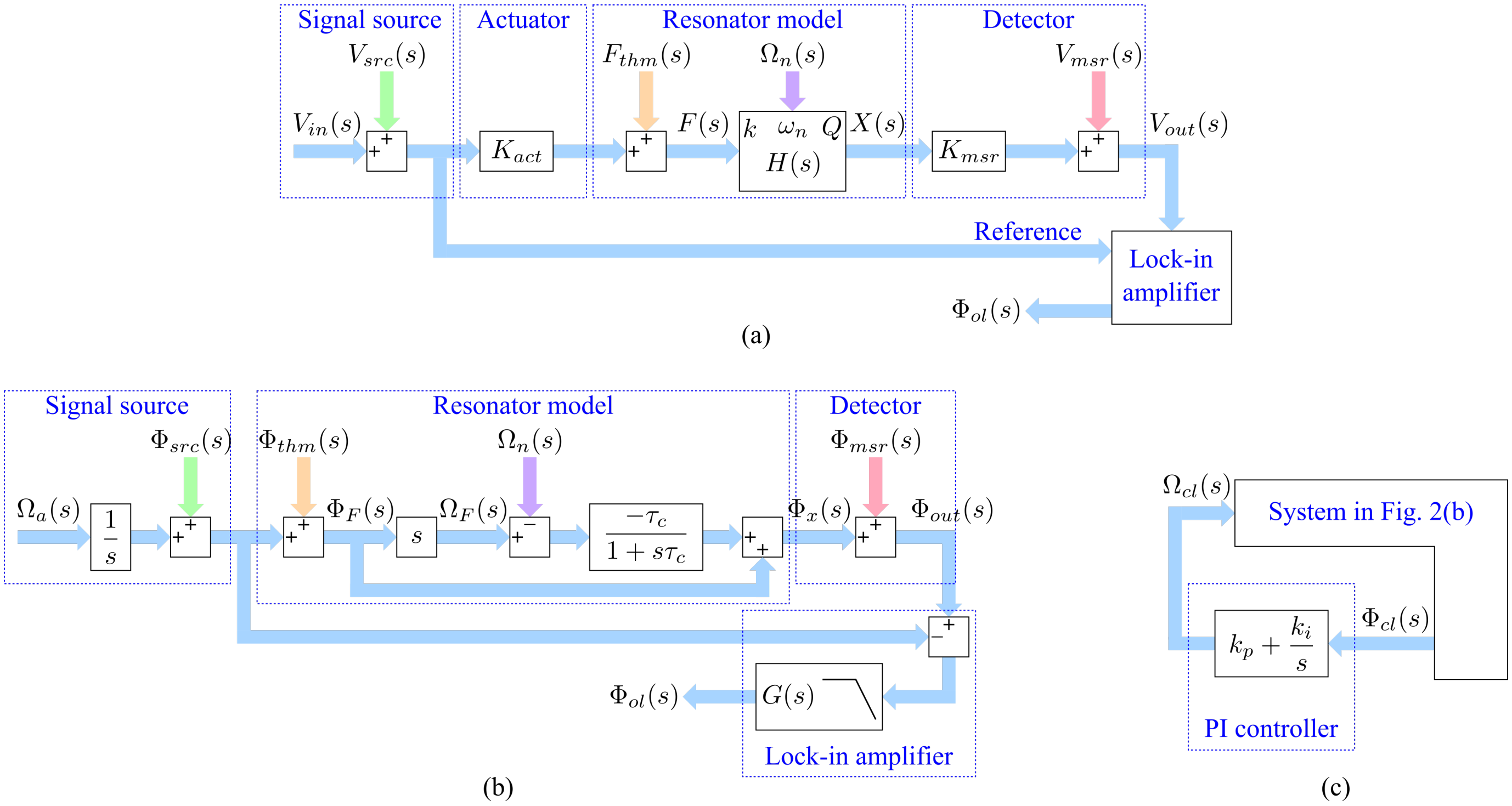

Fig. 1. (a) Model of a resonator with a characterization setup in open loop. In addition to the applied force, the resonance frequency is regarded as an input of the resonator ($\boldsymbol{\Omega_n(s)}$). Inputs $\boldsymbol{V_{src}(s)}$, $\boldsymbol{F_{thm}(s)}$ and $\boldsymbol{V_{msr}(s)}$ model respectively the noise from the signal source, the thermomechanical noise in the resonator, and the noise introduced by the detector. (b) Phase-space model of the same system, linearized around the resonance frequency. The noise inputs in (a) are referred to the phase of the carrier at the corresponding system branch in (b). (c) Phase-space closed-loop system, obtained when a PI controller is connected between the output phase and the input frequency of the system in (b).

For frequency shifts larger than the peak width ($|\Delta\omega| > \omega_n/Q$), (2) does not hold and the phase slope $|\,\mathrm{d}\varphi_0\,/\,\mathrm{d}\omega_a\,|$ reduces, which increases the imprecision of the method. Therefore, for practical applications, a closed-loop approach is usually preferred to increase the measurement range to values beyond the peak width, and reach sensing speeds faster than the open-loop settling time $\tau_c$ [25]. In addition, closed-loop operation ensures that the resonator is actuated at resonance maintaining minimum imprecision, even when resonance shifts occur due to uncontrolled external conditions, like temperature variations.

As discussed in the introduction, finding the limit of mass detection $\delta_m$ involves determining the frequency imprecision $\delta_\omega$. These two quantities are linearly related through:

$$\delta_m = \frac{1}{|R_m|}\delta_\omega, \qquad (3)$$

where $R_m$ is the mass responsivity of the resonance frequency, defined as

$$R_m = \frac{\mathrm{d}\omega_n}{\mathrm{d}m} \approx -\frac{\omega_n}{2m}. \qquad (4)$$

For quantitative analysis, we set the Allan deviation $\sigma_y(\tau)$ [26] equal to the normalized frequency imprecision: $\delta_\omega(\tau)/\omega_n = \sigma_y(\tau)$, where $\tau$ is the gate time, or integration time, used in the evaluation of the Allan deviation. Using (3) and (4), this results in a mass imprecision $\delta_m(\tau) = 2m\sigma_y(\tau)$, which represents the minimum mass that the sensor can resolve, or limit of detection. In Appendix I we show how the Allan deviation can be determined from a noisy displacement signal $x(t)$. To model the Allan deviation resulting from the open-loop and closed-loop configurations of a resonant sensor, we describe below the Laplace-domain representation of both systems.

The open-loop response of a resonant sensor can be analyzed by the block diagram in Fig. 1(a), consisting of an actuator, a resonator and a detector. The system is driven by a signal source producing a harmonic signal $v_{in}(t) = v_0\sin(\omega_a t)$. At the output, a lock-in amplifier determines the phase difference $\varphi_{ol}(t)$ between the driving signal and the measured resonator displacement, that can be used to determine the resonance frequency by using (2) with $\varphi_0 = \varphi_{ol}$. It is worth noting that significant phase shifts might appear in reality from the actuator, the detector and the interconnections. However, these can be assumed constant in time and thus uncorrelated with the sensing events, so that they can be cancelled by subtracting a constant value from $\Phi_{ol}(s)$. Four imprecision sources affecting the system are modelled through different inputs: signal source noise $V_{src}(s)$, thermomechanical noise $F_{thm}(s)$, resonance frequency fluctuations $\Omega_n(s)$ and detector noise $V_{msr}(s)$, whose descriptions can be found in Appendix II. To simplify the analysis, all the signals are considered approximately harmonic, and the system is linearized around its operating point $\omega_a = \omega_n$ [21], [24]. This results in the phase-space system shown in Fig. 1(b), that models the phase and frequency variations over the operating point, and where all components can be represented by additive terms and invariant transfer function blocks. The transfer functions that relate each of the inputs to the output $\varphi_{ol}(t)$ are derived in Appendix II.

The described system can be operated in closed loop by using the output phase $\varphi_{ol}(t)$ to control the frequency of the driving signal $\omega_a(t)$, introducing a proportional-integral (PI) controller. This results in a PLL scheme as shown in Fig. 1(c), in which $\varphi_{ol}(t)$ and $\omega_a(t)$ are renamed as $\varphi_{cl}(t)$ and $\omega_{cl}(t)$ respectively. The PI controller is configured to maintain an output phase $\varphi_{cl} = -\pi/2$ in steady-state, that corresponds to the resonant condition. Within the PLL bandwidth limit, this ensures that the signal $\omega_{cl}(t)$ follows $\omega_n(t)$, such that $\omega_{cl}(t)$ can be used as measure for the resonance frequency. As detailed in Appendix III, this system is stable with a bandwidth $\omega_{pll}$, provided proper lock-in bandwidth and PI controller constants are selected. The transfer functions that relate each of the system inputs to the output frequency $\omega_{cl}(t)$ are also derived in Appendix III.

As outlined in Appendix I, the Allan deviation of the open-loop and closed-loop systems can be evaluated from the outputs $\varphi_{ol}(t)$ and $\omega_{cl}(t)$ respectively. With the transfer functions obtained in Appendices II and III, the Allan deviation $\sigma_y(\tau)$ resulting from the different noise sources included in our model can be determined for both open-loop and closed-loop configurations.

## III. Relation between Open-Loop and Closed-Loop Frequency Imprecision

In this section, we use the transfer functions from Appendices II and III to obtain expressions for the phase and frequency noise, and use those to determine the system's Allan deviation. From these expressions, a methodology is derived to estimate the closed-loop Allan deviation from open-loop measurements.

The open-loop phase noise can be characterized by the two-sided power spectral density (PSD) $S_{\varphi ol}(\omega)$ of the signal $\varphi_{ol}(t)$. This is obtained for any of the four noise sources considered, by using the corresponding transfer function from (A.9)–(A.12) in Appendix II, and using the relation

$$S_{\varphi ol}(\omega) = \left|H_z^{\varphi ol}(j\omega)\right|^2 S_z(\omega), \tag{5}$$

where $z$ represents the subscript of the relevant noise source. The equivalent frequency noise can be obtained by time differentiation of the phase, which results in a PSD

$$S_{\omega ol}(\omega) = \omega^2 S_{\varphi ol}(\omega). \tag{6}$$

The same can be done to obtain the closed-loop phase noise PSD $S_{\varphi cl}(\omega)$ and frequency noise PSD $S_{\omega cl}(\omega)$ for the different noise sources, by using the corresponding transfer function from (A.13)–(A.16) in Appendix III. With these equations, the relation between the phase and frequency noise PSDs in open-loop and closed-loop is found for each noise source. For instance, for thermomechanical noise (A.9) and (A.13) lead to the following relation between the open-loop phase PSD and the closed-loop frequency PSD:

$$\frac{S_{\omega cl}(\omega)}{S_{\varphi ol}(\omega)} = \frac{1+\tau_c^2\omega^2}{\tau_c^2\left(1+\frac{\omega^2}{\omega_{pll}^2}\right)}, \tag{7}$$

where it is assumed that $G(s) \approx 1$ for $\omega < \omega_{pll}$. Using (A.10), (A.14) for the resonance frequency fluctuations, (A.11), (A.15) for the noise from the detector, or (A.12), (A.16) for the noise from the signal source, the same expression as in (7) is obtained. By applying (6), an equivalent expression that relates the frequency noise in open loop to the frequency noise in closed loop is reached:

$$\frac{S_{\omega cl}(\omega)}{S_{\omega ol}(\omega)} = \frac{1+\tau_c^2\omega^2}{\tau_c^2\omega^2\left(1+\frac{\omega^2}{\omega_{pll}^2}\right)}. \tag{8}$$

A graphical representation of (7) and (8) can be found in Fig. 2, where $\omega_c = 1/\tau_c$. This result on resonant sensors driven by a PLL resembles Leeson's effect in direct-feedback oscillators for timing applications [27].

As seen, we can distinguish two frequency ranges to be analyzed: $\omega_c \ll \omega \ll \omega_{pll}$ and $\omega \ll \omega_c$. In the former, the relation

$$S_{\omega cl}(\omega) = S_{\omega ol}(\omega) \tag{9}$$

holds, while in the latter we have

$$S_{\omega cl}(\omega) = \omega_c^2 S_{\varphi ol}(\omega). \tag{10}$$

The range $\omega > \omega_{pll}$ lacks practical interest, as there the PLL cannot track resonance frequency shifts without significant attenuation, see (A.14) in Appendix III. Interestingly, for frequencies $\omega \ll \omega_c$ it is seen from Fig. 2(b) that the open-loop frequency noise is lower than the closed-loop frequency noise. This can be qualitatively understood from the fact that the integral term of the PI controller $k_i/s$ dominates over the proportional term $k_p$ for low frequencies. As seen in Fig. 1(c), the integral part amplifies the phase noise at low frequencies and feeds it back as frequency noise. As a result, the integral term in the PLL configuration, that is needed to ensure that the system operates near the point of minimum imprecision, on the other hand degrades the frequency imprecision with respect to the open-loop system.

The Allan deviation of an arbitrary signal $z$ with carrier frequency $\omega_a$ can be obtained from the PSD of its frequency noise $S_{\omega z}(\omega)$ through [28]

$$\sigma_{yz}^2(\tau) = \int_{-\infty}^{+\infty} \frac{S_{\omega z}(\omega)}{\pi\omega_a^2}\frac{\sin^4(\tau\omega/2)}{(\tau\omega/2)^2}\,\mathrm{d}\omega\,. \tag{11}$$

In our case, $\omega_a = \omega_n$ is used. It is difficult to evaluate this integral analytically, however one can estimate it by considering that the weighing function $\sin^4(\tau\omega/2)/(\tau\omega/2)^2$ is a peak shaped function with its maximum at $\tau\omega = 2.33$ (i.e. $\omega_{peak} = 2.33/\tau$) and a full width at half maximum (FWHM) of $2.51/\tau$. This implies that the Allan deviation will be dominated by the value of $S_{\omega z}(\omega)$ around $\omega_{peak} = 2.33/\tau$. It follows that, for $2.33\tau_c \gg \tau \gg 2.33/\omega_{pll}$, the range $\omega_c \ll \omega \ll \omega_{pll}$ of $S_{\omega z}(\omega)$ is dominant, and therefore, to a good approximation, (9) can be substituted in (11). The implication is that the open-loop Allan deviation $\sigma_{yol}$ is approximately equal to the closed-loop Allan deviation $\sigma_{ycl}$:

$$\sigma_{ycl} \approx \sigma_{yol}, \quad \text{for} \quad 2.33\tau_c \gg \tau \gg \frac{2.33}{\omega_{pll}}. \tag{12}$$

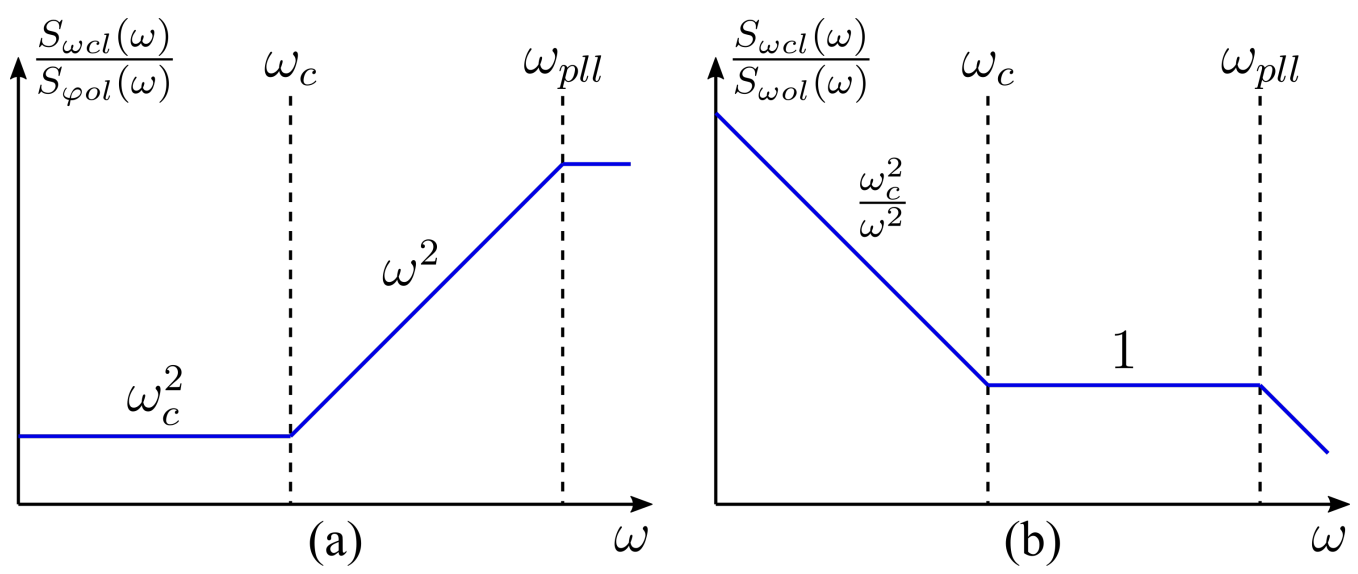

Fig. 2. Asymptotic behaviour in log-log scale of the conversion (a) from open-loop phase noise to closed-loop frequency noise and (b) from open-loop frequency noise to closed-loop frequency noise. These plots result from (7) and (8) respectively. This behaviour is independent of the noise source considered.

By the same reasoning, the closed-loop Allan deviation for $\tau \gg 2.33\tau_c$ can be obtained by combining (10) and (11). Using $\omega_c/\omega_n = 1/(2Q)$, this results in:

$$\sigma_{ycl}^2 \approx \int_{-\infty}^{+\infty} \frac{S_{\varphi ol}(\omega)}{8\pi Q^2} \frac{\sin^4(\tau\omega/2)}{(\tau\omega/2)^2}\, \mathrm{d}\omega\,, \quad \text{for} \quad \tau \gg 2.33\tau_c. \tag{13}$$

The evaluation of the integral requires knowing the frequency dependence of $S_{\varphi ol}(\omega)$ [28].

With the goal of evaluating the closed-loop frequency imprecision from an open-loop measurement, we now derive an equation that translates the measured $\varphi_{ol}(t)$, into the equivalent frequency perturbations $\omega_{cl}(t)$ that the system would experience if operated in closed loop. For that, we need to enforce the PSD relation in (7), which for the particular range of interest $\omega \ll \omega_{pll}$ is

$$\frac{S_{\omega cl}(\omega)}{S_{\varphi ol}(\omega)} = \omega^2 + \omega_c^2. \tag{14}$$

Since $S_{\omega cl}(\omega) = |\Omega_{cl}(j\omega)|^2$, it follows that

$$\frac{\Omega_{cl}(s)}{\Omega_{ol}(s)} = s + \omega_c \tag{15}$$

meets (14). This expression translates into the time domain as

$$\omega_{cl}(t) = \omega_n + \varphi_{ol}(t) * \mathscr{L}^{-1}\{s + \omega_c\}, \tag{16}$$

where $*$ indicates convolution and $\mathscr{L}^{-1}\{\cdot\}$ indicates the inverse Laplace transform. A constant term $\omega_n$ has been reintroduced, that had disappeared in the linearization of the phase-space model. The solution of the last equation gives

$$\omega_{cl}(t) = \omega_n + \frac{\mathrm{d}\varphi_{ol}(t)}{\mathrm{d}t} + \omega_c\varphi_{ol}(t). \tag{17}$$

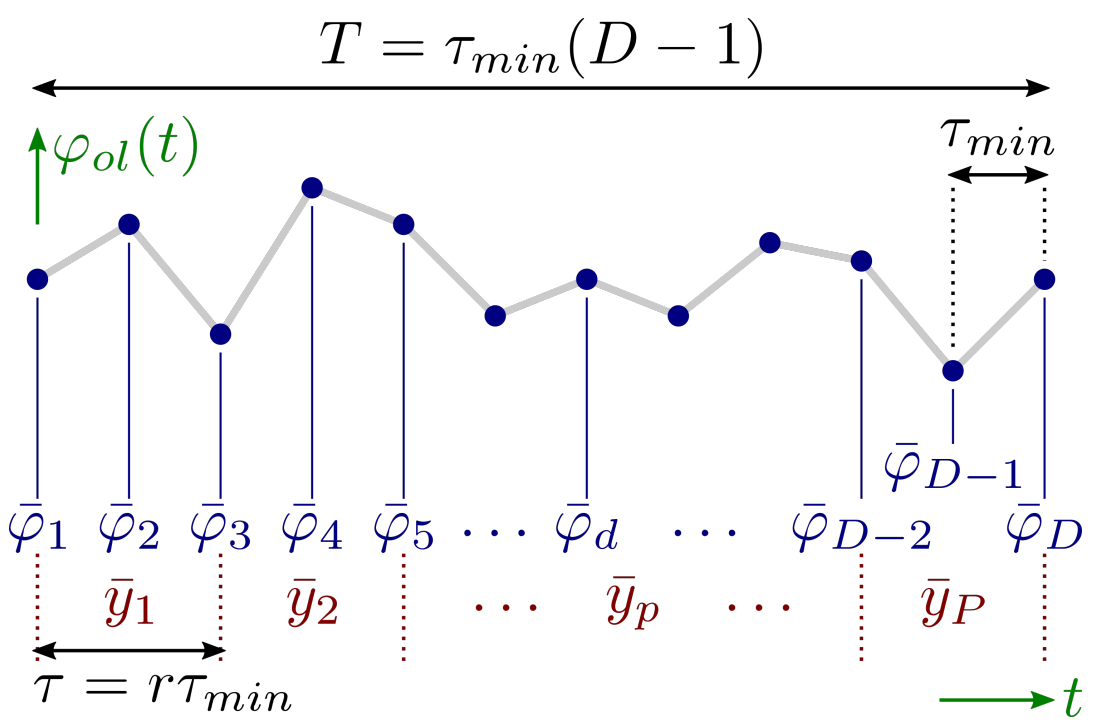

Fig. 3. Graphical representation of a set of phase data points recorded in open loop. An index $d$ identifies each measured point. Time intervals of constant duration are defined and identified with an index $p$. The average fractional frequency is evaluated inside each interval for the calculation of the Allan deviation. The gate time, $\tau$, is defined by the duration of the intervals, $r\tau_{min}$. As an example, $r = 2$ is shown in the figure.

A procedure to perform the conversion dictated by (17) on experimental data of discrete nature is described in Appendix IV. The derivations culminate in expression (A.31), which provides a general formula to obtain the closed-loop Allan deviation $\sigma_{ycl}(\tau)$ from a set of measurements of the resonator's phase in open loop $\varphi_{ol}(t)$, recorded over a significant time span for which the driving frequency stays close to the resonance frequency.

## IV. Method Description

Based on our analysis, we propose an experimental method to evaluate the closed-loop Allan deviation from an open-loop measurement, consisting in the following steps:

1) Implement an open-loop characterization setup for the resonator, consisting of a signal source, an actuator, a detector, and a lock-in amplifier as shown in Fig. 1(a). Determine $\omega_n$ and set the signal source to generate a harmonic signal at $\omega_n$.
2) Record the phase values provided by the lock-in amplifier under constant conditions. As depicted in Fig. 3, the recorded values are labelled as $\varphi_d$, with $d = 1 \ldots D$. From these data, the Allan deviation can be evaluated for gate times $\tau = r\tau_{min}$, where $\tau_{min}$ is the inverse of the sampling rate and $r$ takes integer values. The maximum gate time will be $\tau_{max} = T/\eta$, where $T$ is the recording time and $\eta$ is the minimum number of samples needed for a proper evaluation of the Allan deviation ($M$ in (A.5)), e.g. $\eta = 100$. The bandwidth of the lock-in amplifier must be $\omega_h \gg 2.33/\tau_{min}$.
3) Evaluate the deviation of the recorded phase values from the initial value $\varphi_1$. For large deviations caused by resonance frequency drifts, the method is not valid, and further measures need to be taken to ensure constant experimental conditions. As seen in Appendix V, a rule-of-thumb conservative limit can be set as $\max(\varphi_d - \varphi_1) < 5.7°$ .
4) For each $\tau = r\tau_{min}$, define an index $p = 1 \ldots P$, with $P$ the integer part of $(D-1)/r$. Apply (A.23) and (A.28) over the recorded $\varphi_d$. The results obtained are approximations for the closed-loop Allan deviation valid for $\tau \ll 2.33\tau_c$ and $\tau \gg 2.33\tau_c$ respectively. Using these results, evaluate (A.31) to obtain the closed-loop Allan deviation valid for all $\tau \in [\tau_{min}, \tau_{max}]$.

After having derived this method, it will be validated experimentally in the next section.

## V. Experimental Validation

In order to test the described open-to-closed-loop transformation, a silicon AFM cantilever has been used. Fig. 4(b) shows a top-view picture of this device, which has a length of $245\,\mu m$ and a width of $55\,\mu m$. As depicted in Fig. 4(a), the experiments were carried out in a vacuum chamber at $0.01\,\mathrm{mbar}$. The cantilever was mounted on a piezo-actuator and the measurement of the vibrations was performed by a laser Doppler vibrometer. The signal source, lock-in and PI controller functions were implemented digitally by a Zurich UFHLI lock-in amplifier. As a first characterization, the amplitude and phase of the displacement around the fundamental resonance frequency were measured, and are shown in Fig. 4(c, d). Fitting this response to the transfer function in (1), the resonance frequency and quality factors were calculated, giving $f_n = 165\,\mathrm{kHz}$ and $Q = 6500$. The phase data shows a shift of $\approx 87°$ extrinsic to the resonator, as a result of the phase shifts introduced by the piezo-actuator, the vibrometer and the connection cables.

A PLL as described in Section II was implemented with constants $k_p = 814\,\mathrm{rad\,s^{-1}}$ and $k_i = 65\,135\,\mathrm{rad^2\,s^{-2}}$, that ensured the stability of the loop and established a PLL bandwidth $f_{pll} = 130\,\mathrm{Hz}$ ($1/f_{pll} = 7.7\,\mathrm{ms}$). The phase to be tracked can be set to any value in a digitally implemented PLL, by just subtracting a constant value from the lock-in filter output. In our case, the phase set-point was selected as the resonance value of $-90°$, after correcting for the $87°$ shift introduced by the equipment. The lock-in filter bandwidth was set to $f_h = 10\,\mathrm{kHz}$ with filter order $\alpha = 4$. The signal $\omega_{cl}(t)$, that corresponds to $\Omega_{cl}(s)$ in Fig. 1(c), was recorded with a sampling frequency of $f_s = 24.47\,\mathrm{kHz}$ over a period of 15 minutes and it is shown in Fig. 5(a). A slow drift is observed that can be attributed to uncontrolled experimental conditions, e.g. pressure or temperature, that affect the resonance frequency. Importantly, the closed-loop phase signal $\varphi_{cl}(t)$, whose deviation from the phase set-point represents the tracking error, was centred at zero and showed no drift during the experiment. The Allan deviation was evaluated over $\omega_{cl}(t)$ for different values of $\tau$. The result of this calculation is denoted as $\sigma_{ycl}$ and it is shown in Fig. 5(c). Next, a second experiment was performed in open-loop configuration. In this case, the PI controller was removed and the driving frequency was fixed to the resonance frequency. All other experimental parameters were identical to those in the closed-loop experiment. The recorded signal $\varphi_{ol}(t)$, that corresponds to $\Phi_{ol}(s)$ in Fig. 1(b), is shown in Fig. 5(b). The maximum deviation of the measured $\varphi_{ol}(t)$ from the initial value over the recorded period was $2.5°$. This ensures the validity of the measurement according to the criterion set in Appendix V. Over these data, the procedure described in Appendix IV, using (A.31), was carried out to determine the closed-loop Allan deviation. The result is an estimate of the closed-loop Allan deviation obtained from the open-loop measurement and is denoted as $\sigma_{ycl\leftarrow ol}$. In addition, (A.23) and (A.28) were employed to evaluate the approximations for the closed-loop Allan deviation for short gate times, $\sigma_{ycl\leftarrow ol}|_{\tau\ll 2.33\tau_c}$, and long gate times, $\sigma_{ycl\leftarrow ol}|_{\tau\gg 2.33\tau_c}$, which are also plotted in Fig. 5(c). The intrinsic open-loop Allan deviation $\sigma_{yol}$ is also represented by the light red solid line in Fig. 5(c), as (12) was used in its derivation.

Fig. 5(c) shows the experimental closed-loop fractional frequency Allan deviation (blue solid) and compares it to the Allan deviation as determined from the open-loop experiment by expressions (A.31), (A.23) and (A.28). A good agreement between $\sigma_{ycl}$ and $\sigma_{ycl\leftarrow ol}$ can be observed, providing evidence for the usefulness of the presented method to obtain the imprecision of practical resonant sensors operated in closed loop from open-loop measurements. As seen, formulas (A.23) and (A.28) are good approximations in their validity ranges, separated by $2.33\tau_c$, whereas equation (A.31) is accurate over the full range. Furthermore, as the gate time $\tau$ approaches $2.33/\omega_{pll}$, differences between $\sigma_{ycl}$ and $\sigma_{ycl\leftarrow ol}$ are observed, of at most a factor 2 at the shortest gate times. This can be explained by the smooth cut-off of the PLL transfer function, that starts attenuating the noise (and also the signal) in the closed-loop system for frequencies near and above $\omega_{pll}$. The open-loop estimate of the closed-loop performance is unaffected by this PLL-related artifact, and will only be affected by the lock-in filter cut-off as $\omega$ approaches $\omega_h$ (lock-in filter bandwidth). Therefore, $\sigma_{ycl\leftarrow ol}$ might be regarded as a 'cleaner' estimate of the intrinsic sensor imprecision in closed loop. From a practical point of view, benchmarking a resonator through $\sigma_{ycl\leftarrow ol}$ for a minimum gate time $\tau_{min}$ requires a lock-in amplifier with bandwidth $\omega_h \gg 2.33/\tau_{min}$. However, the direct evaluation of $\sigma_{ycl}$ involves the more demanding requirement of implementing a PLL with bandwidth $\omega_{pll} \gg 2.33/\tau_{min}$.

## VI. Discussion: Validity of the Method

The conversion from open-loop phase noise to closed-loop frequency noise dictated by (17), on which our method relies, is valid for the noise sources included in Fig. 1. These cover all the possible nodes of the system diagram in which noise might be introduced. In addition, this noise relation is equivalent to the Leeson's formula found for direct feedback oscillators [27]. The proposed method is therefore general and can be applied for different types of open-loop and closed-loop systems irrespective of the dominant noise source, provided that the dominant noise source is present in both systems. In this sense, for the PLL analyzed in this work, the method would not be valid if the PI controller, which is present in the PLL but not during the open-loop characterization, introduces noise higher than that of the other system parts. In that case, $\sigma_{ycl}$ would be higher than the predicted $\sigma_{ycl\leftarrow ol}$. Fig. 5(c) shows that this is not the case in our experiments.

If a direct feedback oscillators is considered as closed-loop system (instead of a PLL), care must be taken with the noise introduced by the amplifier in the feedback loop. In such case, the open-loop characterization scheme must include the amplifier at the resonator's output, in order to capture its effects on the closed-loop imprecision predicted by our method.

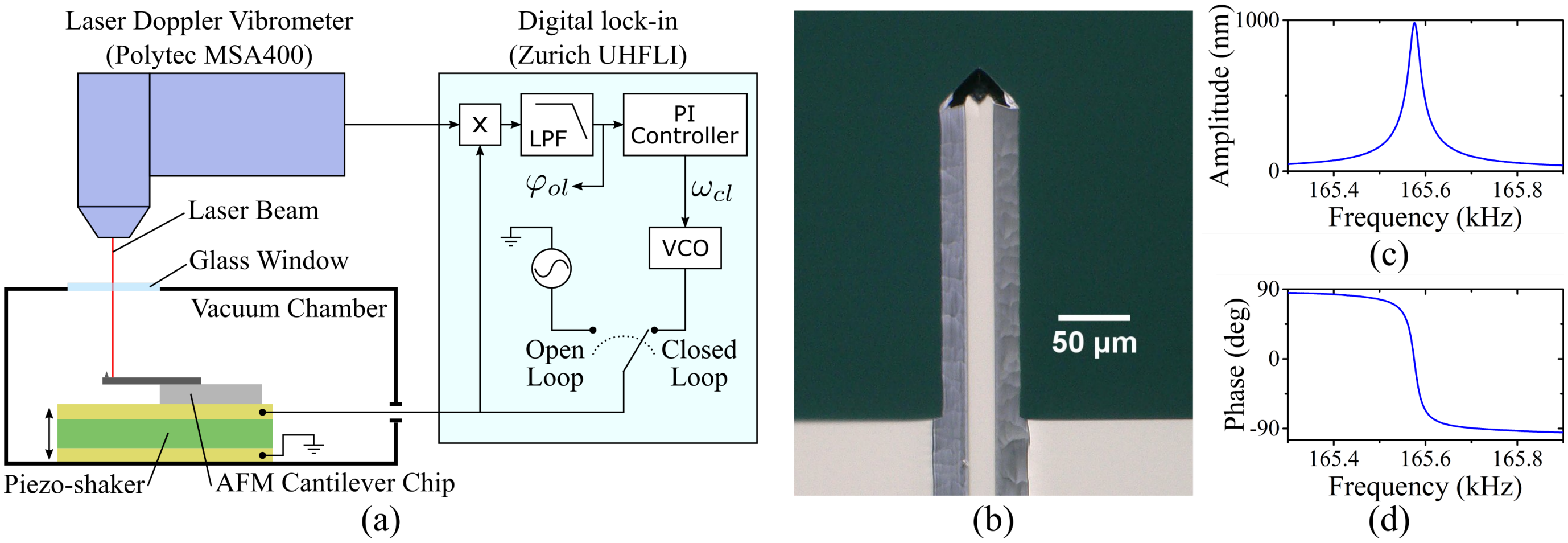

Fig. 4. (a) Schematic of the experimental setup. The two relevant output signals, $\omega_{cl}$ for the closed-loop experiment and $\varphi_{ol}$ for the open-loop experiment, are labelled. (b) Top-view picture of the AFM cantilever used in the experiments. (c) Amplitude and (d) phase of the displacement around the fundamental resonance, for an actuation voltage of 40 mV.

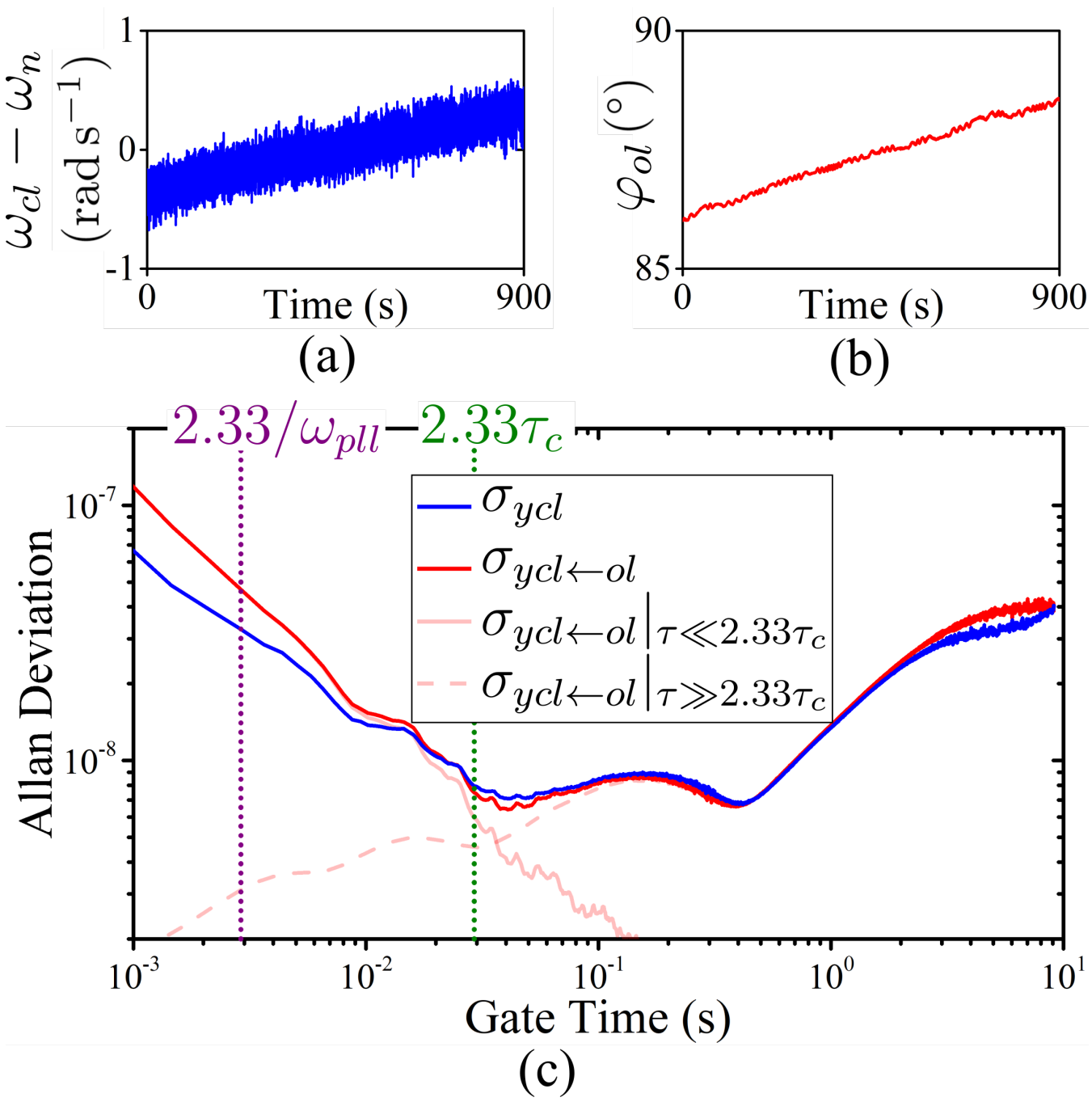

Fig. 5. Raw data recorded in the (a) closed-loop experiment and (b) open-loop experiment. (c) The blue solid line represents the closed-loop Allan deviation $\sigma_{ycl}$ obtained directly from the closed-loop experiment as a function of the gate time $\tau$. The dark red solid line represents the closed-loop Allan deviation obtained from the open-loop experiment $\sigma_{ycl\leftarrow ol}$, using (A.31). The light red solid and dashed lines represent respectively the results provided by the approximated formulas (A.23) and (A.28), applied over the data from the open-loop experiment. The light red solid line in turn represents the open loop Allan deviation $\sigma_{yol}$. Dotted vertical lines indicate the gates time that sets the validity region for the PLL system operation, $\tau > 2.33/\omega_{pll} = 2.9\,\mathrm{ms}$, and the gate time that separates the validity regions for both approximations mentioned, $2.33\tau_c = 29\,\mathrm{ms}$.

We note that the validity of the presented method depends on the assumption made in the evaluation of the integral in (11), which is reasonable unless the frequency noise $S_{\omega z}(\omega)$ varies very strongly with frequency. Another important consideration is that our method assumes that the open-loop characterization is performed with actuation at the resonance frequency so that (2) holds. Nevertheless, drifts experienced by the resonance frequency may cause a large detuning for long experiments, which can compromise the validity of the method. In Appendix V, it is shown that by ensuring a phase drift lower than 5.7° during the open-loop experiment, (2) remains approximately valid with a maximum error of 1% in the slope of the resonator's phase curve. This is confirmed by the fact that the phase recorded during our open-loop experiment shows a drift of 2.5°, despite of what the method remains valid as shown by the results in Fig. 5(c).

## VII. Conclusion

In this work, the relation between the noise observed in resonant sensors operated in open loop and closed loop is investigated. While closed-loop operation allows for higher measurement speed and range, it also entails poorer precision with respect to open-loop operation, due to the change in the system dynamics introduced by the feedback loop.

Based on the analysis of both open-loop and closed-loop resonant sensors, we have derived expressions to estimate the closed-loop Allan deviation based on open-loop measurements. The procedure has been successfully validated using experiments on an AFM cantilever. The presented method is beneficial for fast estimation of the intrinsic precision and benchmarking of resonant sensors, by excluding the effects and efforts related to the implementation of a particular closed-loop configuration.

## Acknowledgment

The authors thank Professors A. Demir and L. G. Villanueva for useful discussions.

## Appendix I
## Determination of the Allan deviation

To analyse the frequency imprecision $\delta_\omega$ from a certain (noisy) displacement signal $x(t)$, we determine the Allan deviation $\sigma_y$ in the following way [28]. Let $x(t)$ be a signal consisting of a harmonic carrier with superimposed noise. If the noise power is much smaller than the carrier power, we can write the noise in terms of the amplitude and phase components, $a_x(t)$ and $\varphi_x(t)$ respectively:

$$x(t) = x_0(1 + a_x(t))\cos(\omega_a t + \varphi_0 + \varphi_x(t)). \tag{A.1}$$

Then, the instantaneous phase is $\omega_a t + \varphi_0 + \varphi_x(t)$ and the instantaneous frequency is its time derivative,

$$\omega_x(t) = \omega_a + \frac{\mathrm{d}\varphi_x(t)}{\mathrm{d}t}. \tag{A.2}$$

The fractional instantaneous frequency is defined by normalizing $\omega_x(t)$ by the carrier frequency:

$$y_x(t) = \frac{\omega_x(t)}{\omega_a} = 1 + \frac{1}{\omega_a}\frac{\mathrm{d}\varphi_x(t)}{\mathrm{d}t}. \tag{A.3}$$

We can now time average $y_x(t)$ over consecutive periods with averaging time $\tau$:

$$\bar{y}_{x,k} = \frac{1}{\tau}\int_{t_k}^{t_{k+1}} y_x(t)\,\mathrm{d}t\,, \quad \text{with} \quad t_{k+1} = t_k + \tau. \tag{A.4}$$

For a large number of periods $M$, the Allan deviation $\sigma_y(\tau)$ of the signal $x(t)$ is defined by the relation

$$\sigma_y(\tau) = \sqrt{\frac{1}{M}\sum_{k=0}^{M-1}\frac{(\bar{y}_{x,k+1} - \bar{y}_{x,k})^2}{2}}. \tag{A.5}$$

The Allan deviation can be interpreted as an estimator for the imprecision of a signal's fractional frequency $\delta_y(\tau) = \delta_\omega(\tau)/\omega_a$ evaluated at a certain gate time $\tau$.

## Appendix II
## Open-Loop Configuration

Fig. 1(a) shows the block diagram that describes the open-loop response of a resonator. A signal source is needed to generate a driving voltage in the form $v_{in}(t) = v_0\cos(\omega_a t)$. The noise of the signal source is indicated by $V_{src}(s)$. A linear actuation mechanism (e.g. piezoelectric, electrostatic or optical), translates the input voltage into force with a proportionality constant $K_{act}$. The resonator is modelled by the transfer function $H(s)$, see (1), in which the resonance frequency is left as a variable. The input resonance frequency $\Omega_n(s)$ accounts for the time-dependent resonance frequency changes due to variations in the sensed parameter (e.g. mass), and also random fluctuations that limit the precision [29]. The additive force $F_{thm}(s)$ models thermomechanical noise, but can also include external forces like those due to unwanted vibrations. At the resonator output, a detector converts the displacement into the voltage signal $V_{out}(s)$, with a proportionality constant $K_{msr}$, and adds measurement noise modelled by $V_{msr}(s)$. Finally, a lock-in amplifier determines the phase of $V_{out}(s)$ with respect to $V_{in}(s)$. The obtained phase, $\varphi_{ol}(t)$, can be translated into an equivalent instantaneous open-loop frequency, $y_{ol}(t)$, by using (A.3). Then, the Allan deviation of $y_{ol}(t)$, $\sigma_{yol}(\tau)$, can be obtained using (A.5).

In order to understand the open-loop response of the resonator to small changes in the resonance frequency, the model in Fig. 1(a) must be linearized. Note that, even if the driving voltage is harmonic and noiseless with frequency $\omega_a$ and the resonator conditions are kept constant, the instantaneous frequency of the excitation force, $\omega_F(t)$, has a random component introduced by the thermomechanical noise $F_{thm}(s)$. For a resonator steadily driven at its resonance frequency ($\Omega_a(s) = \omega_n$) and assuming small noise, $\Omega_F(s) \approx \omega_n$ can be assumed; therefore, the phase difference between displacement and force on the resonator can be obtained in a similar way to (2):

$$\varphi_x - \varphi_F \approx -\frac{\pi}{2} - \tau_c(\omega_F - \omega_n), \tag{A.6}$$

where $\tau_c = 1/\omega_c = 2Q/\omega_n$ is the characteristic time, or settling time, of the resonator. Note that this equation only reflects the steady state response of the resonator. The dynamic relation between the resonator phase and the excitation frequency is derived in [21] for $\omega_F \approx \omega_n$:

$$\Phi_x(s) - \Phi_F(s) = \frac{-\tau_c}{1 + s\tau_c}(\Omega_F(s) - \Omega_n(s)). \tag{A.7}$$

The response obtained shows a single pole at $s = -\omega_c$, indicating that a time of the order of $\tau_c$ is needed to reach steady state. A schematic block diagram of the resulting linearized system of differential equations is shown in Fig. 1(b). This form of the resonator model is usually known as phase-space model, since the variables involved are the phases and frequencies of the different signals, assumed approximately harmonic. Conversion from phase to frequency variables and back can be done respectively by multiplication and division by $s$ in accordance with (A.2) and the properties of the Laplace transform. In the phase-space model of Fig. 1(b), the inputs $V_{src}(s)$, $F_{thm}(s)$ and $V_{msr}(s)$ have been expressed as phase-referred noise sources. The constants $K_{act}$ and $K_{msr}$ do not appear in the phase-space model as they modify the amplitude of the signals but not their phase. The signal source can be represented by a voltage-controlled oscillator (VCO), whose output frequency is controlled by the input $\Omega_a(s)$. This input will help to extend the analysis to the closed-loop configuration. For the open-loop case, the driving frequency is fixed at the resonance frequency to reach the best precision ($\Omega_a(s) = \omega_n$). The linearized model of the lock-in amplifier determines the difference between the phases of the output signal and the driving signal, and low-pass filters the resulting signal with bandwidth $\omega_h$. This filtering is inherent to the operation of phase detection. It is assumed that the low-pass filter is chosen such that $\omega_c \ll \omega_h \ll \omega_n$. A transfer function with an arbitrary number of poles $\alpha$ (the higher $\alpha$ the sharper the filter roll-off above $\omega_h$) is used to describe the lock-in filter:

$$G(s) = \left(\frac{\omega_h}{s + \omega_h}\right)^\alpha. \tag{A.8}$$

From the phase-space system described in Fig. 1(b), transfer functions can be obtained in a straightforward manner to relate each input to the output $\Phi_{ol}(s)$:

$$H^{\varphi_{ol}}_{\varphi_{thm}}(s) = \frac{\Phi_{ol}(s)}{\Phi_{thm}(s)} = \frac{1}{1+\tau_c s}G(s) \quad (A.9)$$

$$H^{\varphi_{ol}}_{\omega_n}(s) = \frac{\Phi_{ol}(s)}{\Omega_n(s)} = \frac{-\tau_c}{1+\tau_c s}G(s) \quad (A.10)$$

$$H^{\varphi_{ol}}_{\varphi_{msr}}(s) = \frac{\Phi_{ol}(s)}{\Phi_{msr}(s)} = G(s) \quad (A.11)$$

$$H^{\varphi_{ol}}_{\varphi_{src}}(s) = \frac{\Phi_{ol}(s)}{\Phi_{src}(s)} = \frac{-\tau_c s}{1+\tau_c s}G(s) \quad (A.12)$$

Since the system is linear, the different contributions to $\Phi_{ol}(s)$ are additive and the effect of each input can be analyzed separately assuming the other inputs are zero. As seen, (A.10) has its lowest frequency pole at approximately $-\omega_c$, which implies that the resonator in open loop cannot respond to changes in the resonant frequency at rates faster than $\tau_c$, because it takes a number of periods of the order of $Q$ to reach steady state.

## Appendix III
## Closed-Loop Configuration

In Appendix II, a set of equations has been derived to relate the considered noise sources to the phase noise at the system output, assuming the driving frequency and the resonant frequency are constant and equal. In this Appendix, we derive a set of equations equivalent to (A.9)–(A.12) for the closed-loop system.

The PLL system works by continuously evaluating the phase difference between the driving signal and the measured displacement and adjusting the driving frequency such that the resonant condition, which is identified by a phase difference of $-\pi/2$ between $F(s)$ and $X(s)$, is continuously enforced. The open-loop system in Fig. 1(b) takes the driving frequency $\Omega_a$ as input and gives the phase lag introduced by the resonator $\Phi_{ol}$ as output. In the closed-loop system, shown in Fig. 1(c), a controller is introduced to continuously determine the driving frequency as a function of the phase lag. As detailed in [21], this can be done by a proportional-integral (PI) controller. $\Phi_{cl}(s)$ and $\Omega_{cl}(s)$ will be used to denote respectively the output phase and input frequency of the system under closed-loop operation, see Fig. 1(c). The closed-loop frequency imprecision can then be determined by calculating the Allan deviation $\sigma_{ycl}(\tau)$ of the time-domain signal $\omega_{cl}(t)$, defined as the inverse Laplace transform of $\Omega_{cl}(s)$. The obtained Allan deviation defines the fractional frequency imprecision for the closed-loop operation of the resonator, $\delta_{\omega cl}(\tau)/\omega_n = \sigma_{ycl}(\tau)$.

The transfer function of the PI controller is $H_{PI}(s) = k_p + k_i/s$. This results in a total open-loop gain of the system, including the controller that tracks the resonance frequency, of $H^{\varphi_{ol}}_{\omega_n} H_{PI}$. Therefore, the corresponding closed-loop transfer function becomes $H^{\omega_{cl}}_{\omega_n} = -H^{\varphi_{ol}}_{\omega_n} H_{PI}/(1 - H^{\varphi_{ol}}_{\omega_n} H_{PI})$. Choosing the PI constants in such a way that $k_i = k_p/\tau_c$ results in a simplified transfer function for the closed-loop system, by a zero-pole cancellation. As detailed in [21], this choice simplifies the tuning of the PLL bandwidth $\omega_{pll}$, but it does not affect the system behaviour within that bandwidth. Setting the lock-in filter bandwidth as $\omega_h \gg k_p$ gives a PLL bandwidth $\omega_{pll} \approx k_p$. To simplify expressions, $k_p = \omega_{pll}$ will be used in the following. With these definitions, the closed-loop transfer functions relating each of the inputs in Fig. 1(b) to $\Omega_{cl}$ (s) are:

$$H^{\omega_{cl}}_{\varphi_{thm}}(s) = \frac{\Omega_{cl}(s)}{\Phi_{thm}(s)} = \frac{\omega_{pll}}{\tau_c(\omega_{pll}G(s)+s)}G(s) \quad (A.13)$$

$$H^{\omega_{cl}}_{\omega_n}(s) = \frac{\Omega_{cl}(s)}{\Omega_n(s)} = \frac{-\omega_{pll}}{\omega_{pll}G(s)+s}G(s) \quad (A.14)$$

$$H^{\omega_{cl}}_{\varphi_{msr}}(s) = \frac{\Omega_{cl}(s)}{\Phi_{msr}(s)} = \frac{\omega_{pll}(1+\tau_c s)}{\tau_c(\omega_{pll}G(s)+s)}G(s) \quad (A.15)$$

$$H^{\omega_{cl}}_{\varphi_{src}}(s) = \frac{\Omega_{cl}(s)}{\Phi_{src}(s)} = \frac{\omega_{pll}s}{\omega_{pll}G(s)+s}G(s) \quad (A.16)$$

Equation (A.14) shows that the system can indeed track frequency shifts up to a bandwidth of approximately $\omega_{pll}$, provided $\omega_h \gg \omega_{pll}$ (or $G(s) \approx 1$ for $\omega < \omega_{pll}$). Thus, in closed-loop configuration, the system can be set to achieve shorter response times than the characteristic time of the resonator $\tau_c$ by making $\omega_{pll} > \omega_c$.

## Appendix IV
## Evaluation of the Closed-Loop Frequency Imprecision from Open-Loop Measurements

A procedure to evaluate the closed-loop frequency imprecision from open-loop measurements is described. The procedure starts from the time-domain phase recorded at the output of the lock-in amplifier in the open-loop system of Fig. 1(b), $\varphi_{ol}(t)$. As depicted in Fig. 3, we will identify each measured phase value by a subscript $d$:

$$\bar{\varphi}_d = \frac{1}{\tau}\int_{(d-\frac{3}{2})\tau_{min}}^{(d-\frac{1}{2})\tau_{min}} \varphi_{ol}(t)\,\mathrm{d}t\,, \quad (A.17)$$

with $d$ ranging from 1 to $D$. $\tau_{min}$ is the inverse of the sampling rate, that sets the minimum gate time (integration time) $\tau$ for which the Allan deviation can be evaluated. Gate times of multiples of $\tau_{min}$ are possible by using integration intervals of duration $\tau = r\tau_{min}$, with multiplication integer $r$. These successive intervals will be identified by the subscript $p$, and run from $t = (p-1)\tau$ to $t = p\tau$.

As deduced in (9), for $2.33\tau_c \gg \tau \gg 2.33/\omega_{pll}$ the frequency noise in closed loop and open loop are equal, so we can make $\omega_{cl}(t) = \omega_{ol}(t)$. Using the fractional frequency definition in (A.3), we obtain $y_{cl}(t) = y_{ol}(t)$. With $\omega_a = \omega_n$, this results in

$$y_{cl}(t)|_{\tau \ll 2.33\tau_c} = 1 + \frac{1}{\omega_n}\frac{\mathrm{d}\varphi_{ol}(t)}{\mathrm{d}t}. \quad (A.18)$$

For each interval $p$, the average closed-loop average fractional frequency $\bar{y}_{p,cl}$ can be obtained as

$$\bar{y}_{p,cl} = \frac{1}{\tau}\int_{(p-1)\tau}^{p\tau} y_{cl}(t)\,\mathrm{d}t\,, \quad (A.19)$$

with $p = 1 \dots P$. Then, the closed-loop average fractional frequencies inside each interval, $\bar{y}_{p,cl}$, are obtained combining (A.18) and (A.19), giving

$$\bar{y}_{p,cl}|_{\tau \ll 2.33\tau_c} = 1 + \frac{\varphi_{ol}(p\tau) - \varphi_{ol}((p-1)\tau)}{\tau\omega_n}. \quad (A.20)$$

By inspecting Fig. 3, it can be seen that the best estimator for $\varphi_{ol}(p\tau)$ from the available data is $\bar{\varphi}_{1+pr}$, so that (A.20) becomes

$$\bar{y}_{p,cl}|_{\tau\ll 2.33\tau_c} = 1 + \frac{\bar{\varphi}_{1+pr} - \bar{\varphi}_{1+(p-1)r}}{\tau\omega_n}. \quad \text{(A.21)}$$

The closed-loop Allan deviation $\sigma_{ycl}$ can now be obtained by calculating

$$\sigma^2_{ycl}(\tau) = \frac{1}{2(P-1)} \sum_{p=1}^{P-1} (\bar{y}_{p+1,cl} - \bar{y}_{p,cl})^2. \quad \text{(A.22)}$$

By substituting (A.21) in (A.22), the closed-loop Allan deviation can be expressed as a function of the recorded data points:

$$\sigma^2_{ycl}(\tau)|_{\tau\ll 2.33\tau_c} = \frac{1}{2(P-1)(\tau\omega_n)^2} \sum_{p=1}^{P-1} (\bar{\varphi}_{1+(p+1)r} - 2\bar{\varphi}_{1+pr} + \bar{\varphi}_{1+(p-1)r})^2. \quad \text{(A.23)}$$

For $\tau \gg 2.33\tau_c$, the values calculated with (A.21) do not represent the frequency perturbations in closed loop. In this $\tau$ range, the measured phase perturbations in open loop must be translated into frequency perturbations in closed loop in accordance with (10). This is ensured by using the relation

$$\omega_{cl}(t)|_{\tau\gg 2.33\tau_c} = \omega_c \varphi_{ol}(t), \quad \text{(A.24)}$$

which with $\omega_c/\omega_n = 1/(2Q)$ gives

$$y_{cl}(t)|_{\tau\gg 2.33\tau_c} = \frac{\omega_{cl}(t)}{\omega_n} = \frac{\varphi_{ol}(t)}{2Q}. \quad \text{(A.25)}$$

Introducing this expression in (A.19), the closed-loop average fractional frequencies are

$$\bar{y}_{p,cl}|_{\tau\gg 2.33\tau_c} = \frac{1}{\tau} \int_{(p-1)\tau}^{p\tau} \frac{\varphi_{ol}(t)}{2Q}\, \mathrm{d}t\,. \quad \text{(A.26)}$$

Given the discrete nature of the measured data $\bar{\varphi}_d$, the best approximation for this expression is obtained by averaging the recorded phase values inside each interval:

$$\bar{y}_{p,cl}|_{\tau\gg 2.33\tau_c} = \frac{1}{2Qr} \sum_{d=1+(p-1)r}^{pr} \bar{\varphi}_d. \quad \text{(A.27)}$$

Expression (A.22) can be applied over these $\bar{y}_{p,cl}$ to obtain the closed-loop Allan deviation for $\tau \gg \tau_c$. Again, we can express it as a function of the recorded data points $\bar{\varphi}_d$:

$$\sigma^2_{ycl}(\tau)|_{\tau\gg 2.33\tau_c} = \frac{1}{8(P-1)r^2Q^2} \sum_{p=1}^{P-1} \left[ \left( \sum_{d=1+pr}^{(p+1)r} \bar{\varphi}_d \right) - \left( \sum_{d=1+(p-1)r}^{pr} \bar{\varphi}_d \right) \right]^2. \quad \text{(A.28)}$$

It must be noted that (A.23) and (A.28) describe the asymptotic behaviour of the closed-loop Allan deviation for short and long $\tau$. Equation (17) obtained in Section III provides an open-loop to closed-loop transformation valid for the full range of interest $\omega \ll \omega_{pll}$, and can be expressed for the fractional frequency as

$$y_{cl}(t) = \frac{\omega_{cl}(t)}{\omega_n} = 1 + \frac{1}{\omega_n}\frac{\mathrm{d}\varphi_{ol}(t)}{\mathrm{d}t} + \frac{\varphi_{ol}(t)}{2Q}. \quad \text{(A.29)}$$

Resorting to the previous analysis, it can be seen that this equation comprisses a first term equivalent to (A.18), and a second term equivalent to (A.25). Therefore, (A.29) results in the following expression for the average fractional frequencies:

$$\bar{y}_{p,cl} = \bar{y}_{p,cl}|_{\tau\ll 2.33\tau_c} + \bar{y}_{p,cl}|_{\tau\gg 2.33\tau_c}. \quad \text{(A.30)}$$

By introducing these values in (A.22), the closed-loop Allan deviation over the full range of $\tau$ longer than the PLL integration time (all $\tau \gg 2.33/\omega_{pll}$) can be written in terms of the measured open-loop phase data:

$$\sigma^2_{ycl}(\tau) = \sigma^2_{ycl}(\tau)|_{\tau\ll 2.33\tau_c} + \sigma^2_{ycl}(\tau)|_{\tau\gg 2.33\tau_c} + \frac{1}{2r(P-1)Q\tau\omega_n} \sum_{p=1}^{P-1} \left[ \left(\bar{\varphi}_{1+(p+1)r} - 2\bar{\varphi}_{1+pr} + \bar{\varphi}_{1+(p-1)r}\right) \times \left( \sum_{d=1+pr}^{(p+1)r} \bar{\varphi}_d - \sum_{d=1+(p-1)r}^{pr} \bar{\varphi}_d \right) \right] \quad \text{(A.31)}$$

Equations (A.23) and (A.28) can be used to evaluate the first two terms.

## Appendix V
## Effect of Resonance Frequency Drifts During Open-Loop Measurements

If in the open-loop measurements the difference between the driving frequency and resonance frequency becomes too large, the linear phase-frequency relation in (2) breaks down. Thus, for the validity of our method, random perturbations and drifts of the resonance frequency must remain small, so that $\omega_a \approx \omega_n$ for the duration of the data recording. The maximum frequency shift tolerated can be expressed as a fraction of the peak width ($\omega_n/Q$):

$$|\omega_a - \omega_n|_{max} < \epsilon\frac{\omega_n}{Q}, \quad \text{(A.32)}$$

where $\epsilon$ can be regarded as the safety coefficient. For small values of $\epsilon$, we can apply (2) to obtain that, within this range, the maximum deviation of the resonator phase from $-\pi/2$ (the value at resonance) is $\pm\epsilon$. As an example, it can be calculated from (1) that making $\epsilon = 0.1$ ensures that the slope of the phase curve of $H(s)$ deviates a maximum of 1% from $-\tau_c$ (the value at resonace). This means a maximum phase difference of $\pm\epsilon = \pm 0.1\,\text{rad} = \pm 5.7°$. According to this criterion of 'maximum slope deviation of 1%', if the driving frequency is set to the resonance frequency at the beginning of the open-loop experiment, the recorded data points $\bar{\varphi}_d$ must not deviate more than $\pm 5.7°$ from the initial value $\bar{\varphi}_1$.